\documentclass[12pt]{elsart}
\usepackage{graphicx}
\usepackage{amsmath}
\usepackage{amsfonts}
\usepackage{amssymb}

\begin{document}

\journal{Phys. Lett. A}

\begin{frontmatter}
\title{Molecules in clusters: the case of planar LiBeBCNOF built from a
triangular form LiOB and a linear four-center species FBeCN}
\author[CTChem]{G. Forte}
\author[CTChem]{A. Grassi}
\author[CTChem]{G. M. Lombardo}
\author[CTPhys]{G. G. N. Angilella\thanksref{corr}}
\author[Oxford,AntwerpPhys]{N. H. March}
\author[CTPhys]{R. Pucci}

\address[CTChem]{Dipartimento di Scienze Chimiche,\\
Facolt\`a di Farmacia, Universit\`a di Catania,\\
Viale A. Doria, 6, I-95126 Catania, Italy}
\address[CTPhys]{Dipartimento di Fisica e Astronomia, Universit\`a di
Catania,\\ and CNISM, UdR Catania, and INFN, Sez. Catania,\\ 64, Via S. Sofia,
I-95123 Catania, Italy}
\address[AntwerpPhys]{Department of Physics, University of Antwerp,\\
Groenenborgerlaan 171, B-2020 Antwerp, Belgium}
\address[Oxford]{Oxford University, Oxford, UK}
\thanks[corr]{Corresponding author. E-mail: {\tt giuseppe.angilella@ct.infn.it}.}

\begin{abstract}
Kr\"uger some years ago proposed a cluster LiBeBCNOF, now called periodane. His
ground-state isomer proposal has recently been refined by Bera \emph{et al.}
using DFT. Here, we take the approach of molecules in such a cluster as starting
point. We first study therefore the triangular molecule LiOB by coupled cluster
theory (CCSD) and thereby specify accurately its equilibrium geometry in free
space. The second fragment we consider is FBeCN, but treated now by restricted
Hartree-Fock (RHF) theory. This four-center species is found to be linear, and
the bond lengths are obtained from both RHF and CCSD calculations. Finally, we
bring these two entities together and find that while LiOB remains largely
intact, FBeCN becomes bent by the interaction with LiOB. Hartree-Fock and CCSD
theories then predict precisely the same lowest isomer found by Bera \emph{et
al.} solely on the basis of DFT.

\end{abstract}

\end{frontmatter}

\section{Introduction}

Though traditional chemical thinking in which `atoms in molecules' was a prime
focus goes back many decades (see \emph{e.g.} the early work of Moffitt
\cite{Moffitt:51}), more recently the idea has been championed most notably by
Bader and coworkers \cite{Bader:94}. Here, we have been motivated by the
proposal of Kr\"uger \cite{Krueger:06} on LiBeBCNOF, termed periodane, to study a
`molecules in clusters' approach to this species. This cluster, treated
subsequently, in a more refined quantum chemical manner than in
\cite{Krueger:06}, by Bera \emph{et al.} \cite{Bera:07}, is found to be
essentially planar. One molecular grouping which then seemed apparent was LiOB,
with the strong bond being B--O. This left Be, C, and N, which would form a
radical and would be spin-compensated in the ground state by adding F.

Below, calculations are reported on the ground-state isomer of periodane, by (a)
both restricted Hartree-Fock (RHF) and coupled cluster (CCSD) theory, and (b)
for LiOB by the coupled cluster singles and doubles (CCSD) methods. Thereby we
can make a quantitative comparison with Bera \emph{et al.} \cite{Bera:07} of
both bond lengths and angles in the final geometry of the predicted lowest
isomer from DFT.

\section{Quantum chemical methodology}

Several isomers of periodane were considered and, for each of them, a geometry
optimization has been performed by using a 6-311(d) standard basis set
\cite{McLean:80} for all the atoms at Hartree-Fock level. The geometry of the
most stable isomer, thermodynamically speaking, was further optimized at coupled
cluster with single and double excitations (CCSD) level
\cite{Cizek:69,Purvis:82,Scuseria:88}. All the calculations have been performed
by using the G03 package \cite{Frisch:04}.

\begin{figure}[b]
\centering
\includegraphics[width=0.8\columnwidth]{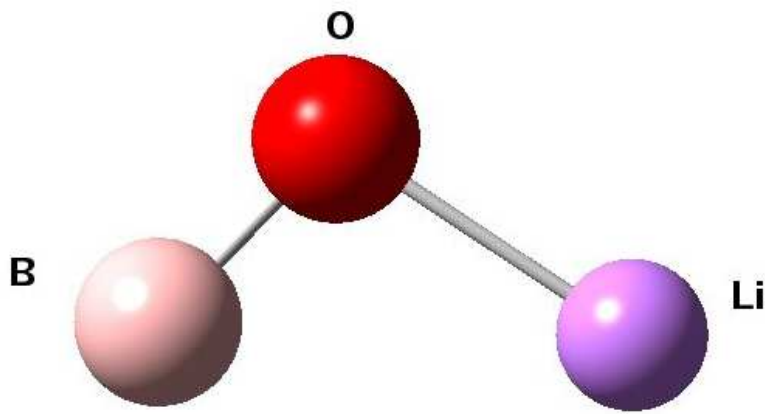}
\caption{(Color online) Shows predicted geometry of molecule LiOB as obtained
from a CCSD calculation. The bond lengths, the angle, and the energy are given
in Tab.~\ref{tab:params}.}
\label{fig:LiOB}
\end{figure}

\begin{figure}[t]
\centering
\includegraphics[width=0.4\columnwidth]{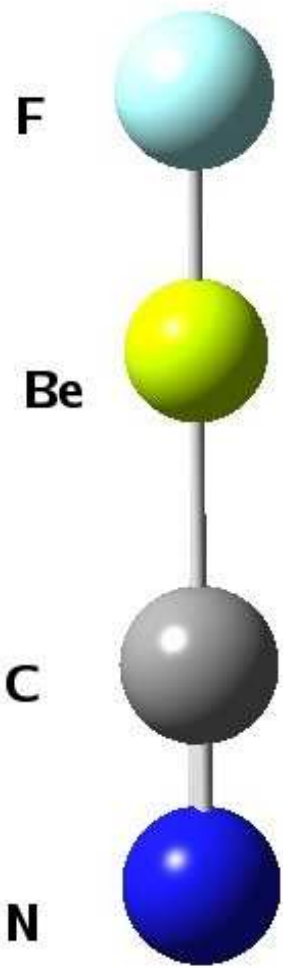}
\caption{(Color online) Depicts linear geometry found for FBeCN four-center molecule.
Structural parameters are listed in Tab.~\ref{tab:params}.}
\label{fig:FBeCN}
\end{figure}

\begin{figure}[t]
\centering
\includegraphics[width=0.8\columnwidth]{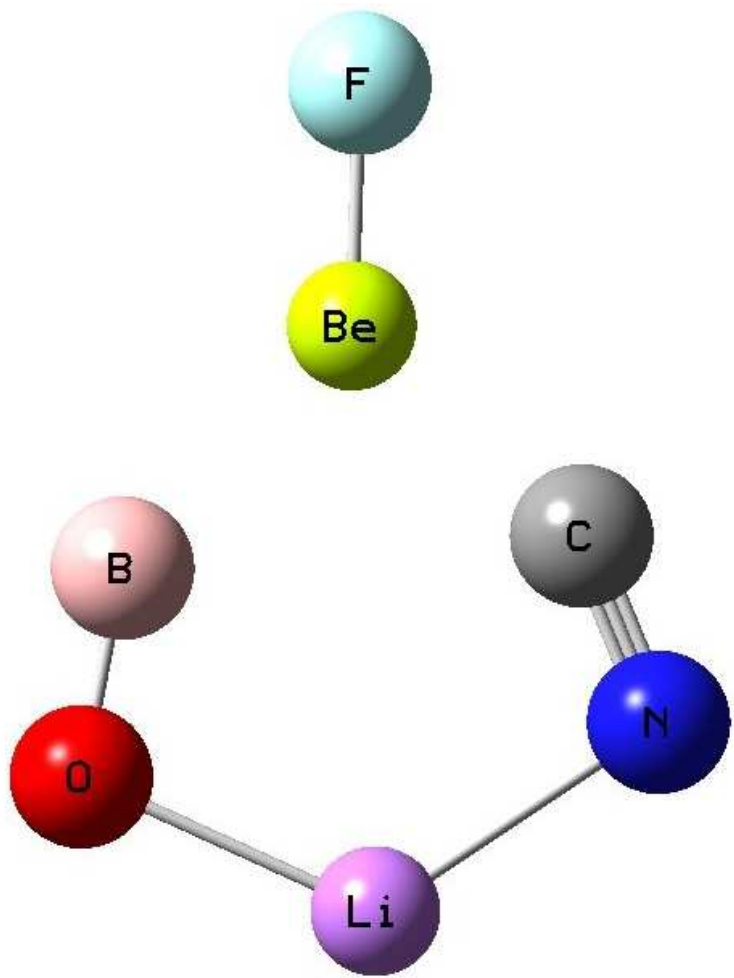}
\caption{(Color online) Shows predicted lowest isomer for periodane from RHF and CCSD theories
(with 6-311(d) basis set). Bond lengths and angles (Tab.~\ref{tab:params}) agree
excellently with DFT predictions \cite{Bera:07}. The two strongest bonds are
seen to be B--O and N--C, with Be--F also short.}
\label{fig:LiBeBCNOF}
\end{figure}

\begin{table*}[t]
\caption{Bond lengths (in \AA), angles (in degrees), and energies (in Hartree)
of the lowest isomers of LiOB (Fig.~\ref{fig:LiOB}), FBeCN
(Fig.~\ref{fig:FBeCN}), and periodane (LiBeBCNOF, Fig.~\ref{fig:LiBeBCNOF}),
within RHF and CCSD methods.}
\label{tab:params}
\begin{tabular}{l|r|rr|rr}
\hline\hline
       & \multicolumn{1}{c|}{LiOB} & \multicolumn{2}{c|}{FBeCN} &
       \multicolumn{2}{c}{LiBeBCNOF} \\
       & \multicolumn{1}{c|}{CCSD} & \multicolumn{1}{c}{RHF} &
       \multicolumn{1}{c|}{CCSD} & \multicolumn{1}{c}{RHF} &
       \multicolumn{1}{c}{CCSD} \\
\hline
Distances (\AA)  & & & & & \\
O--B & 1.26 & & & 1.21 & 1.23 \\
Li--O & 1.75 & & & 1.86 & 1.88 \\
C--N & & 1.13 & 1.17 & 1.14 & 1.17 \\
Be--C & & 1.67 & 1.66 & 1.83 & 1.81 \\
F--Be & & 1.36 & 1.37 & 1.39 & 1.40 \\
N--Li & & & & 2.00 & 1.98 \\
B--Be & & & & 1.96 & 1.94 \\
\hline
Angles ($^\circ$)  & & & & & \\
B--O--Li & 101 & & & 105.3 & 102.9 \\
Be--C--N & & & & 155.3 & 154.9 \\
C--N--Li & & & & 101.6 & 100.1 \\
N--Li--O & & & & 120.2 & 123.0 \\
O--B--Be & & & & 146.9 & 147.3 \\
B--Be--C & & & & 90.6 & 91.8 \\
\hline
Energy (Hartree)  & & & & & \\
 & $-107.423$ & $-206.545$ & $-207.176$ & $-313.705$ & $-315.362$ \\
\hline\hline
\end{tabular}
\end{table*}

\section{Molecules in clusters}

The triatomic molecule LiOB has then been studied specifically and accurately by
CCSD (see \emph{e.g.} Ref.~\cite{Bartlett:95} for a review). With the B--O
strong bond, the geometry predicted by CCSD is shown in Fig.~\ref{fig:LiOB}, the
ground-state energy being $-107.422807$~Hartree. With the DFT functional of
\cite{Bera:07}, a similar geometry was found with a lower energy of about
0.25~Hartree: It is not clear to us that the DFT variational value lies above
the exact ground-state energy because of approximations in the energy functional
that are needed to date.

For the spin-compensated four-center molecule FBeCN, a linear structure was
obtained as shown in Fig.~\ref{fig:FBeCN}, where the structural parameters are
recorded in Tab.~\ref{tab:params}.

Figure~\ref{fig:LiBeBCNOF} shows schematically the way the two isolated
molecules, with the individual geometries cited above, are somewhat modified as
they are brought together into what we predict, as do Bera \emph{et al.}
\cite{Bera:07} by purely DFT, as the lowest isomer of periodane. The four-center
molecule is clearly distorted from linearity, the bond lengths and angles being
recorded in Table~\ref{tab:params}. The change in the triatomic LiOB is seen to
be much smaller than in the four-center case.
Table~\ref{tab:eigens} reports the sum of the RHF eigenvalues for the occupied
orbitals for (i) FBeCN as in Fig.~\ref{fig:FBeCN}, and for (ii) isolated FBeCN,
but with all constituent atoms held rigid at the HF geometry in
Fig.~\ref{fig:LiBeBCNOF} for periodane. The HF eigenvalue sums are seen to be
quite close for linear and bent geometries and hence somewhat subtle corrections
to Walsh's rules \cite{Walsh:53} discussed in \cite{March:81} are required to
determine the relative stability between linear and bent forms of FBeCN.

\begin{table}[t]
\caption{Reports the sum of the RHF eigenvalues (in Hartree) for the occupied
orbitals for (i) FBeCN as in Fig.~\ref{fig:FBeCN}, and for (ii) isolated FBeCN,
but with all constituent atoms held rigid at the HF geometry in
Fig.~\ref{fig:LiBeBCNOF} for periodane.}
\label{tab:eigens}
\centering
\begin{tabular}{rr|rr}
\hline\hline
\multicolumn{2}{c|}{(i) FBeCN} & \multicolumn{2}{c}{(ii) FBeCN} \\
\hline
HF & CCSD & HF & CCSD \\
$-129.992$ & $-129.972$ & $-129.640$ & $-129.642$ \\
\hline\hline
\end{tabular}
\end{table}

\begin{table}[t]
\caption{Reports energy differences (in Hartree) between the given clusters and
their constituent isolated atoms.}
\label{tab:endiff}
\centering
\begin{tabular}{l|rr}
\hline\hline
 & HF & CCSD \\
 \hline
LiOB & $-0.306$ & $-0.414$ \\
FBeCN & $-0.489$ & $-0.643$\\
\hline\hline
\end{tabular}
\end{table}

\section{Summary}

The structure of the lowest isomer as predicted by Hartree-Fock and CCSD theory
is shown in Fig.~\ref{fig:LiBeBCNOF}. As stated above, our `molecules in
clusters' approach has led to an identical structure reached by Bera \emph{et
al.} \cite{Bera:07} on the basis of DFT alone. We have argued that it is useful
to be viewed as having building blocks of (a) the bent triatomic molecule LiOB,
and (b) the linear four-center molecule formed from BeCN plus F. Both these
molecules are stable against dissociation into their isolated neutral atoms (cf.
Tab.~\ref{tab:endiff}, reporting the difference between the energies of the
clusters LiOB amd FBeCN and their isolated constituent atoms, respectively).

Of course, as pointed out very specifically in \cite{Bera:07}, it is never
possible to exclude the possibility of a (slightly) lower energy isomer than
that shown in Fig.~\ref{fig:LiBeBCNOF}. Nevertheless, we believe that the
present picture of bringing together two molecules LiOB and the linear
four-center system BeCN plus F is a favourable way of approaching the final
structure of the lowest isomer.

\ack

NHM made his contribution to the present study during a visit to the University
of Catania. He wishes to thank Professors Porto and Pucci for very generous
hospitality.

\bibliographystyle{mprsty}
\bibliography{a,b,c,d,e,f,g,h,i,j,k,l,m,n,o,p,q,r,s,t,u,v,w,x,y,z,zzproceedings,Angilella}

\end{document}